\documentclass{article}
\usepackage[a4paper, portrait, margin=1.1811in]{geometry}
\usepackage[english]{babel}
\usepackage[utf8]{inputenc}
\usepackage[T1]{fontenc}
\usepackage{helvet}
\usepackage{etoolbox}
\usepackage{graphicx}
\usepackage{titlesec}
\usepackage{caption}
\usepackage{booktabs}
\usepackage{xcolor} 
\usepackage[colorlinks, citecolor=cyan]{hyperref}
\usepackage{caption}
\usepackage{amsmath,bm}
\usepackage{float}

\graphicspath{ {./images/} }
\usepackage{scrextend}
\usepackage{fancyhdr}
\usepackage{graphicx}
\newcounter{lemma}

\newcounter{theorem}

\fancypagestyle{plain}{
	\fancyhf{}

}

\makeatletter
\patchcmd{\@maketitle}{\LARGE \@title}{\fontsize{16}{19.2}\selectfont\@title}{}{}
\makeatother

\usepackage{authblk}

\newsavebox\affbox
\author[1]{\textbf{Ali Shirazi}}
\author[2]{\textbf{Fereshteh Sadeghi Naieni Fard}}
\affil[1]{ Bank of America, New York, Director in Quantitative Strategies and Data Group 
}
\affil[2]{ Department of Information Science, Graduate Student, University of North Texas
}

\titlespacing\section{0pt}{12pt plus 4pt minus 2pt}{0pt plus 2pt minus 2pt}
\titlespacing\subsection{12pt}{12pt plus 4pt minus 2pt}{0pt plus 2pt minus 2pt}
\titlespacing\subsubsection{12pt}{12pt plus 4pt minus 2pt}{0pt plus 2pt minus 2pt}

\titleformat{\section}{\normalfont\fontsize{10}{15}\bfseries}{\thesection.}{1em}{}
\titleformat{\subsection}{\normalfont\fontsize{10}{15}\bfseries}{\thesubsection.}{1em}{}
\titleformat{\subsubsection}{\normalfont\fontsize{10}{15}\bfseries}{\thesubsubsection.}{1em}{}

\titleformat{\author}{\normalfont\fontsize{10}{15}\bfseries}{\thesection}{1em}{}

\title{\textbf{\huge Financial Hedging and Risk Compression}\\
	A journey from linear regression to neural network}
\date{April 2023}    

\begin{document}

\pagestyle{headings}	
\newpage
\setcounter{page}{1}
\renewcommand{\thepage}{\arabic{page}}

\captionsetup[figure]{labelfont={bf},labelformat={default},labelsep=period,name={Figure }}	\captionsetup[table]{labelfont={bf},labelformat={default},labelsep=period,name={Table }}
\setlength{\parskip}{0.5em}
	
\maketitle
	
\noindent\rule{15cm}{0.5pt}
	\begin{abstract}
Finding the hedge ratios for a portfolio and risk compression is the same mathematical problem. Traditionally, regression is used for this purpose. However, regression has its own limitations. For example, in a regression model, we can’t use highly correlated independent variables due to multicollinearity issue and instability in the results. A regression model cannot also consider the cost of hedging in the hedge ratios estimation.  We have introduced several methods that  address the linear regression limitation while achieving better performance. These models, in general, fall into two categories: Regularization Techniques and Common Factor Analyses. In regularization techniques, we minimize the variance of hedged portfolio profit and loss (PnL) and the hedge ratio sizes, which helps reduce the cost of hedging. The regularization techniques methods could also consider the cost of hedging as a function of   the cost of funding, market condition, and liquidity. In common factor analyses, we first map variables into common factors and then find the hedge ratios so that the hedged portfolio doesn't have any exposure to the factors. We can use linear or nonlinear factors construction. We are introducing a modified beta variational autoencoder that constructs common factors nonlinearly to compute hedges. Finally, we introduce a comparison method and generate numerical results for an example.   \\ \\
		\let\thefootnote\relax\footnotetext{
			\small $^{*}$\textbf{Corresponding author.} \textit{
				\textit{E-mail address: \color{cyan}alishirazi@gmail.com}}\\
		                                  }
		\textbf{\textit{Keywords}}: \textit{financial hedging; cost of hedging; neural network; regularization;common factors}
	\end{abstract}
\noindent\rule{15cm}{0.4pt}

\section{Introduction}
Market risk management is a very important part of any financial business in both buy and sell sides. Assume we have a portfolio of one or more securities. We are interested to hedge the portfolio against market risk by adding, long or short positions, of additional securities to minimize the portfolio Profit and Loss (PnL) given a market movement. In addition, we could monitor our risk exposure by measuring the portfolio sensitivity to risk factors movements.
As we will show later, both of these problems are related and mathematically the same. The challenge is to find the proper hedging instruments and compute the hedge ratios. 
Traditionally linear regressions are used to compute the hedge ratios or for risk compression. However, there are limitations in using regression especially when there are multiple hedge instruments that are highly correlated. Or when we need to select a subset of hedge instruments from a number of candidate instruments considering not only minimizing risk but also hedge ratio stability and cost.

\section{Hedge Ratio vs. Risk Compression}
Assume we have a portfolio of multiple securities that we need to manage against market risk. First, assume we want to add additional "n" securities as  hedging instruments to our portfolio. For any market movement, we can break down the PnL of the whole portfolio into two buckets:
\begin{eqnarray}
	PnL_{total}(t)= PnL_{nhedged}(t) - PnL_{hedging}(t)
	\label{Eq1}
\end{eqnarray}

If we have done a really good job selecting the right instruments and the right amount of each hedging instrument then $ PnL_{total}$ should be zero for any given market movement that generates $PnL_{unhedged}$ and $ PnL_{Hedging}$. But in reality $ PnL_{total}$ can't be zero but rather a small value either positive or negative. In other words, we are trying to minimize the variance of $ PnL_{total}$. 
 If we also present $PnL_{total}(t)$   by $\epsilon(t) $ and we show $PnL_{unhedged}(t) $ as $Y(t) $ then we have:

\begin{eqnarray}
	PnL_{total}(t)= PnL_{unhedged}(t) - \sum_{i=1}^{n} \beta_i H_i(t) 
	\label{Eq2}
\end{eqnarray}
\begin{eqnarray}
	Y(t) =  \sum_{i=1}^{n} \beta_i H_i(t)  + \epsilon(t)
	\label{Eq3}
\end{eqnarray}
In this equation, $H_i(t)$ represents the movement in hedge instrument "i" while $\beta_i$ is the corresponding hedge ratio. Mathematically we are trying to find $\beta_i$ values i.e. hedge ratios, so the hedged portfolio $PnL_{total}(t)$ or $\epsilon(t)$  has the lowest variance. This is a regression fitting problem. \\
In risk compression, we are interested in monitoring our portfolio risk. For example, assume our portfolio has exposure to "m" risk factors i.e. $RF_i$ when $i=1,2,..,m$. We can compute the sensitivity of our portfolio PnL to those risk factors:
\begin{eqnarray}
	S_i(t) = \frac{\partial PnL(t)}{\partial x_i}
	\label{Eq4}
\end{eqnarray}
The problem with computing the sensitivity using the partial derivative of PnL with respect to any risk factor is that it does not consider the correlation between risk factors. In addition, in practice,  monitoring the sensitivities over time for all  risk factors are difficult given a large number of risk factors. One solution to these problems is to clone the PnL of the main portfolio by using a handful of financial instruments so that at any given time the PnL of the main portfolio is equal or very close to the one of the cloned portfolio. 
\begin{eqnarray}
	PnL(t)= \sum_{i=1}^{i=k <m} \alpha_i ~RF_i(t)  + \epsilon'(t)
	\label{Eq5}
\end{eqnarray}
As it can be seen, both finding hedging ratio and risk compression are mathematically the same challenges i.e. finding the best hedging instruments and then finding the right hedge ratios.

\section{Regression Limitations}

Traditionally finding hedge ratios (or risk compression) is achieved by using linear regression. However, not all requirements of a regression model  could be met in practice. In particular, hedging instruments are correlated and high correlation in the independent variables creates multicollinearity issue and instability in the  estimated factor loadings i.e. hedge ratios. Mathematically speaking, one or more independent variables could be written as a linear combination of other independent variables and this will cause the correlation matrix of independent variables to not be full rank and inversable. \\
We also assume that hedge instruments are given in the regression.  However, in practice, we need to select a few hedging instruments from a pool of candidate instruments first and then find the hedge ratios. It is also desirable to hedge a portfolio against market risk with minimum hedging cost. The cost of hedging is a function of how much each hedging instrument is required as well as the unit cost of the hedging instruments. A regression model can't consider the additional requirements and these limitations.

\section{Alternative Methods}
As explained in the previous section, one of the requirements of regression is to use relatively independent variables that are not highly correlated to avoid multicollinearity. Multicollinearity  could be detected using different statistical tests e.g. Variance Inflation Factor (VIF) \cite{VIF}. \\
In practice, it is not always possible to use independent hedge instruments. In general, there are two approaches to dealing with highly correlated hedge instruments:
\begin{itemize}
    \item Regularization Techniques.
    \item Common Factor Analyses.
\end{itemize}
In the following sections, we explain each method and how it could be employed to estimate our hedge estimation. 

\section{Regularization Techniques}
Assume we want to hedge the PnL of a portfolio where we present the PnL time series as $Y(t)$ using N number of hedging instruments presented by $X_i$ time series. The independent variables could have  high correlations where if used in a regression equation, we will have a multicollinearity issue.\\
In addition, one of the concerns in statistical learning models is overfitting. We are using historical data as a proxy for what might happen in the future. We would like to learn from the past but if we create a model that completely fits the historical data then it will not perform well in the future when new and unseen data comes.  One of the techniques used to overcome this issue is to calibrate and estimate the parameters to not only achieve a good fit to historical data but also restrain the coefficients to become too big. When we have multicollinearity a number of coefficients could become big and unstable.  Therefore we could use regularization to overcome the multicollinearity as well as overfitting. In general, we estimate the regression coefficient using the following equation:
\begin{eqnarray}
	\tilde{\mathbf{\beta}}=\mathop  {argmax}\limits_{\beta } \{  \sum_{i=1}^{N} (y_i +\beta_0- \sum_{j=1}^{P} x_{ij}\beta_j )^2+\lambda\sum_{j=1}^{P}\mid \beta_j \mid ^P\} \cite{Hastie}
	\label{Eq6}
\end{eqnarray}
The following figure shows the Contours of the constant value of  $\mid \beta_j \mid ^P$  for given values of P. If P is selected as 1 the equation turns into  least absolute shrinkage and selection operator; also Lasso or LASSO. It was originally introduced in geophysics \cite{Santosa}, and later by Robert Tibshirani  who coined the term. \cite{Tibshirani}. 

\begin{figure}[H]
	\centering
	\includegraphics[width=0.8\textwidth]{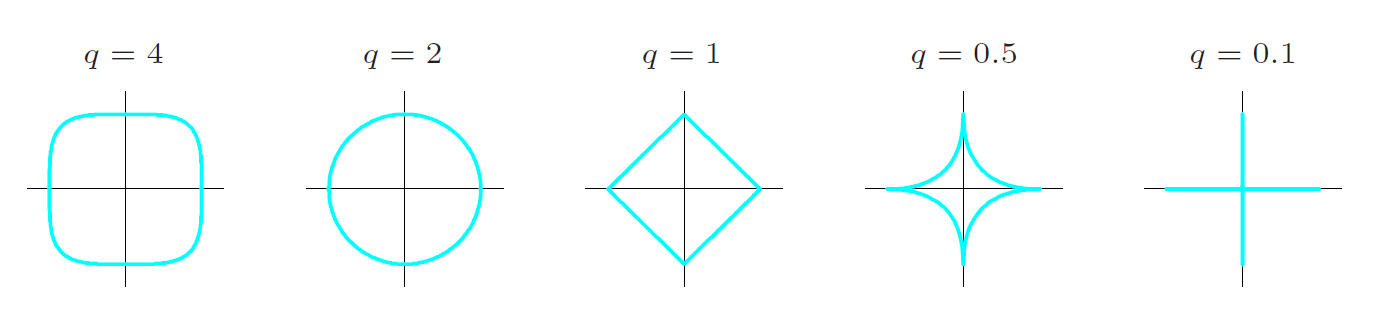}
	\caption{Contours of constant value of  $\mid \beta_j \mid ^P$  for given values of q. \cite{Hastie} }
	\label{fig1}
\end{figure}
If P is equal to 2 then the equation turns into Ridge Regression also known as Tikhonov regularization \cite{Hoerl}. The theory was first introduced by Hoerl and Kennard in 1970 in their Technometrics papers “RIDGE regressions: biased estimation of nonorthogonal problems” and “RIDGE regressions: applications in nonorthogonal problems”.\\
Parameter $\lambda$ controls the amount of shrinkage in factor loadings i.e. $\beta$ values. There is usually a trade-off in the amount of hedge instruments  that should  be purchased versus better market risk control. The PnL of a hedged portfolio could be written as:
\begin{eqnarray}
	PnL_{Hedged}(t) = PnL_{Hedging~Cost}(t) + PnL_{Market}(t)
	\label{Eq7}
\end{eqnarray}
As the above equation shows the hedged portfolio PnL has two components:
\begin{enumerate}
  \item $PnL_{Hedging~Cost}(t):$ This is usually a deterministic negative PnL associated with the cost of hedging. It is desirable to minimize this cost. 
  \item $PnL_{Market}(t):$ This is a stochastic value with the expected value of zero which could be either positive or negative due to the market movement. If the expected value is not zero then it is referred to as alpha. The goal is to minimize the variance of this term. 
\end{enumerate}
Typically we can't achieve items 1 and 2 simultaneously. We rather have to find the optimum solution. This is achieved by either cross-validation or manually changing $\lambda$ to achieve targeted market risk hedges and the number/amount of hedge instruments.\\
Even though there are similarities between Ridge and Lasso but in practice, they could be used in different scenarios:
\begin{itemize}
    \item \textbf{Ridge}: If we know the candidate hedge instrument(s), we can use Ridge regression. This will ensure no hedge ratio is equal to zero and gets eliminated. 
    \item \textbf{Lasso}: If we are not sure about which hedging instrument(s) should be used, we can include all candidate hedging instruments and Lasso might force a number of hedge ratios equal to zero and only keep the best instruments while computing the hedge ratios.  
\end{itemize}
In other words, Lasso could be used as a feature engineering method while ridge will keep all independent variables by  forcing the squared value of variables to be small but not zero values.\\
One of the considerations in finding the best hedges is the cost of hedging. As discussed  before, shrinkage methods in general reduce the cost by applying restrictions on the size of factor loadings. However, if the cost of hedging is not the same for all hedge instruments then it is desirable to consider the unit cost of each hedging instrument in the optimization. This could be achieved using the following equation:
\begin{eqnarray}
\tilde{\mathbf{\beta}}=\mathop  {argmax}\limits_{\beta } \{  \sum_{i=1}^{N} (y_i +\beta_0- \sum_{j=1}^{P} x_{ij}\beta_j )^2+\lambda\sum_{j=1}^{P}\mid \psi_j \beta_j \mid ^P\}  
	\label{Eq8}
\end{eqnarray}
 In this equation, $\psi_j $ is the relative unit cost of $ \beta_j$. Given $\lambda$ is scaling the second term,  $\psi_j $ only needs to be the relative unit cost of $ \beta_j$ instead of the actual unit cost of $ \beta_j$. Please note that the cost of hedging for  securities is not uniformly defined and could change for different financial institutions. In addition, the cost of shorting an instrument i.e. a negative $ \beta_j$ might not be the same as buying  (having a long position) on the same security i.e. a positive $ \beta_j$. Later in this section, we discuss how the unit cost of hedging could be computed.\\
Generally, libraries and packages can't consider the unit cost of hedges ( $\psi_j $ ). The following change of variable could be used to change the equation to be able to use available libraries and packages to solve the optimization problem.
\begin{eqnarray}
\hat{\beta_j}= \psi_j \beta_j
 \label{Eq9}
\end{eqnarray}
\begin{eqnarray}
\hat{x}_{ij}= \frac{x_{ij}}{\psi_j }
 \label{Eq10}
\end{eqnarray}
\begin{eqnarray}
\tilde{\mathbf{\hat{\beta}}}=\mathop {argmax}\limits_{\hat{\beta }} \{  \sum_{i=1}^{N} (y_i +\beta_0- \sum_{j=1}^{P} \hat{x}_{ij} \hat{\beta_j} )^2+\lambda\sum_{j=1}^{P}\mid \hat{ \beta_j} \mid ^P\}  
\label{Eq11}
\end{eqnarray}
In this method, we first adjust each independent variable with its relative unit cost. Then we compute the factor loadings using the adjusted independent variables and the standard algorithm for optimization. The hedge ratios are computed after adjusting back the factor loadings by their unit costs.\\ 
The unit cost of security could be computed as:
\begin{eqnarray}
\psi= r + \frac{E(S_{bid-ask})}{Price_{ask}}  
\label{Eq12}
\end{eqnarray}
\begin{eqnarray}
S_{bid-ask}= Price_{ ask} - Price_{ bid}
\label{Eq13}
\end{eqnarray}
where $r$ is the cost of funding for a unit of the hedging instrument price. r is not only a function of the market but also the creditworthiness of the institution as well as the hedging instrument. For example, the cost of funding could change for different instruments or even the direction of trade i.e. long vs. short.  $E(S_{bid-ask})$  is the expected  bid-ask spread when the hedges are liquidated. We are assuming that we have to pay for the bid-ask spread. Even as a market maker, hedging is a risk-reduction activity and we are not expecting to get the benefit of the bid-ask spread. The expected bid-ask spread is a function of market trading volume and in general liquidity for that security. Selecting a liquid hedge instrument will decrease the cost associated with the bid-ask spread and also lower cost of funding.  The unit in this equation is the cost of a unit of hedging security price.

\section{Common Factor Analyses}
\label{section:FA}
Another way to deal with multicollinearity is to use Common Factor Analyses. The idea is to find common factors between hedge instruments and the main portfolio. And then use the common factors to find the hedge instruments.\\
Assume we want to hedge the PnL of a portfolio where we present the PnL time series as $Y(t)$ using N hedging instruments presented by $X_i$ time series. The independent variables could have  high correlations where if used in a regression equation, we will have a multicollinearity issue.  We first find the time series of N factors  so that  factors are either independent or do not have high correlations between them so that their correlation matrix is full rank. The factors should also be able to explain a majority of the variance of the variables (both main portfolio PnL as well as each hedging instrument).
\begin{eqnarray}
	X_{i}(t) =  \sum_{i=1}^{n} \gamma_i F_i(t)  + \epsilon(t)
	\label{Eq14}
\end{eqnarray}
where $ X_i(t)$ is the value of each hedging instrument and $F_j$ stands for each common factor. please note that $E(\epsilon (t)) = 0$. Please also note that we are using the same number of factors as the hedging instruments. We can write the equation in the matrix form as:
\begin{eqnarray}
	\bm{X_{k \times N}}= \bm{ F_{k \times N}~\gamma_{N \times N}}+ \bm{\epsilon_{k \times N}} 
	\label{Eq15}
\end{eqnarray}
where k is the number of data points and N is the number of variables (both hedging instruments and common factors).
We can write PnL(t) i.e. $Y(t)$ as a function of factors as well:
\begin{eqnarray}
	Y(t) = \sum_{j=1}^{N}\alpha_{ij}  F_j + \delta_i (t)
	\label{Eq16}
\end{eqnarray}

where $E(\delta_i (t)) = 0$. We can write the equation in matrix form:
\begin{eqnarray}
	 \bm{Y_{k \times 1}}= \bm{ F_{k \times N}~ \alpha_{N \times 1}}+ \bm{\delta_{k \times 1}} 
	\label{Eq17}
\end{eqnarray}
If the  hedged PnL has no exposure to any factor and given the factors explain the majority of the variance then it means we are reducing the PnL after hedging:
\begin{eqnarray}
	 \bm{PnL^{unHedged}_{k \times 1}} - \bm{ X_{k \times N}~ \beta_{N \times 1}}=  
\bm{ F_{k \times N}~ \alpha_{N \times 1}}+ \bm{\delta{k \times 1}} - 
(\bm{ F_{k \times N}~\gamma_{N \times 1}}+ \bm{\epsilon_{k \times N}}) \bm{\beta_{N \times 1}}  
	\label{Eq18}
\end{eqnarray}
If we set the coefficients of factors equal to zero we will have:
\begin{eqnarray}
	 \bm{\alpha_{N \times 1}-\gamma_{N \times N}~\beta_{N \times 1}=0_{N \times 1} }
	\label{Eq19}
\end{eqnarray}
\begin{eqnarray}
	 \bm{\beta_{N \times 1}= \gamma_{N \times N}^{-1}~\alpha_{N \times 1}}
	\label{Eq20}
\end{eqnarray}
\begin{eqnarray}
	 \bm{PnL^{Hedged}_{k \times 1}= \delta_{k \times 1} +\epsilon_{k \times N}~\gamma_{N \times N}^{-1}~\alpha_{N \times 1} }
	\label{Eq21}
\end{eqnarray}

The factors could be determined using different methods.  In general, chosen factors should have the following characteristics: 
\begin{itemize}
  \item Factors should be able to explain the majority of variables' variance.
  \item Factors should be either independent or have low correlations between them.
  \item Factors should satisfy regression requirements e.g. stationarity etc.
  \item $\bm{\gamma}$ matrix as define by equation \ref{Eq15} should be inversable. 

\end{itemize}

\vspace{10pt}
 There are different methods to build or select factors. Factors could be selected from multiple factor models e.g. Fama–French three-factor model \cite{ff} for equity. However, there might not be proper multiple factor models for all asset classes that satisfy the requirements.\\
 In general, we could construct factors mathematically. For example Principal Component Analysis, PCA \cite{PCA} is a good method to extract factors. Another candidate for factor construction is to leverage factor analysis. "Factor analysis is a statistical method used to describe variability among observed, correlated variables in terms of a potentially lower number of unobserved variables called factors" \cite{factor}.
\section{Neural Network}
In this section, we describe how neural networks could be leveraged in finding hedge ratios. "Artificial neural networks (ANNs), usually simply called neural networks (NNs) or neural nets, are computing systems inspired by the biological neural networks that constitute animal brains". \cite{nnt}. You can think of a neural network as a function where it gets inputs and generates output(s). The functions are usually  non-linear. You might think to use hedge instruments as inputs and the main portfolio PnL as the output and calibrate the model to mimic the portfolio PnL. 
\begin{figure}[H]
	\centering
	\includegraphics[width=0.8\textwidth]{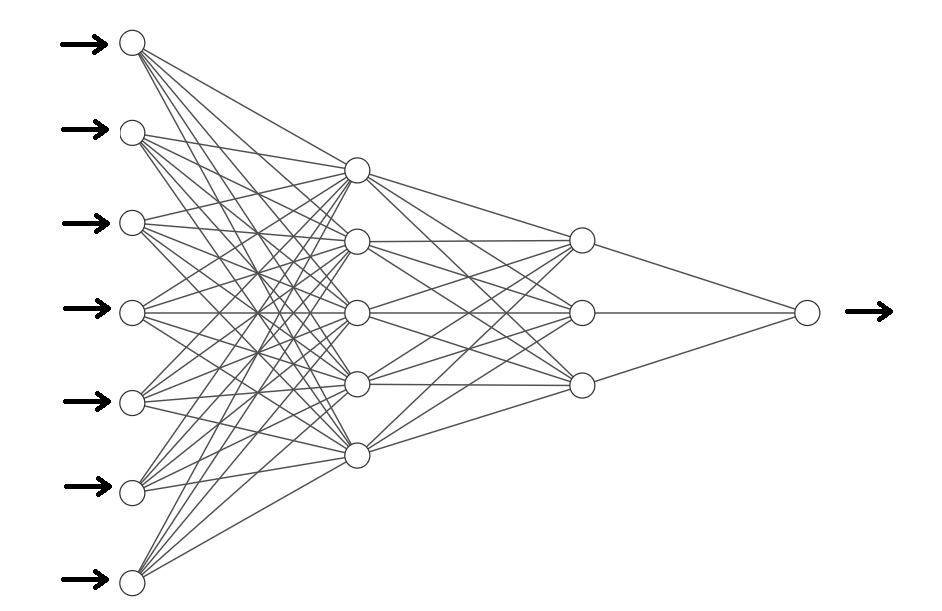}
	\caption{Neural Network where inputs are hedge instruments and output is the portfolio PnL }
	\label{fig2}
\end{figure}
However, the main issue with using a simple Neural Network as shown in Figure \ref{fig2} is that even though it can model PnL using  hedge instruments accurately, the transformation from hedge instrument to PnL is non-linear. Generally, we can't create a portfolio that mimics the non-linear function of instruments. In order to overcome this issue we are designing a particular form of  neural networks to overcome this issue as we explain in the following sections.
\subsection{Autoencoders}

As explained in section \ref{section:FA} we can map hedge instruments into factors and use them to compute hedging ratios. Factors could be constructed either as a linear combination of instruments or a  nonlinear combination of instruments. For example, in PCA we construct factors as a linear combination of instruments under certain restrictions to maximize the variances of principal components. One of the methods to generate factors nonlinearly is to use autoencoders."An autoencoder is a type of artificial neural network used to learn efficient codings of unlabeled data (unsupervised learning). The encoding is validated and refined by attempting to regenerate the input from the encoding. The autoencoder learns a representation (encoding) for a set of data, typically for dimensionality reduction, by training the network to ignore insignificant data (“noise”)." \cite{aencod}\\

\begin{figure}[H]
	\centering
	\includegraphics[width=0.8\textwidth]{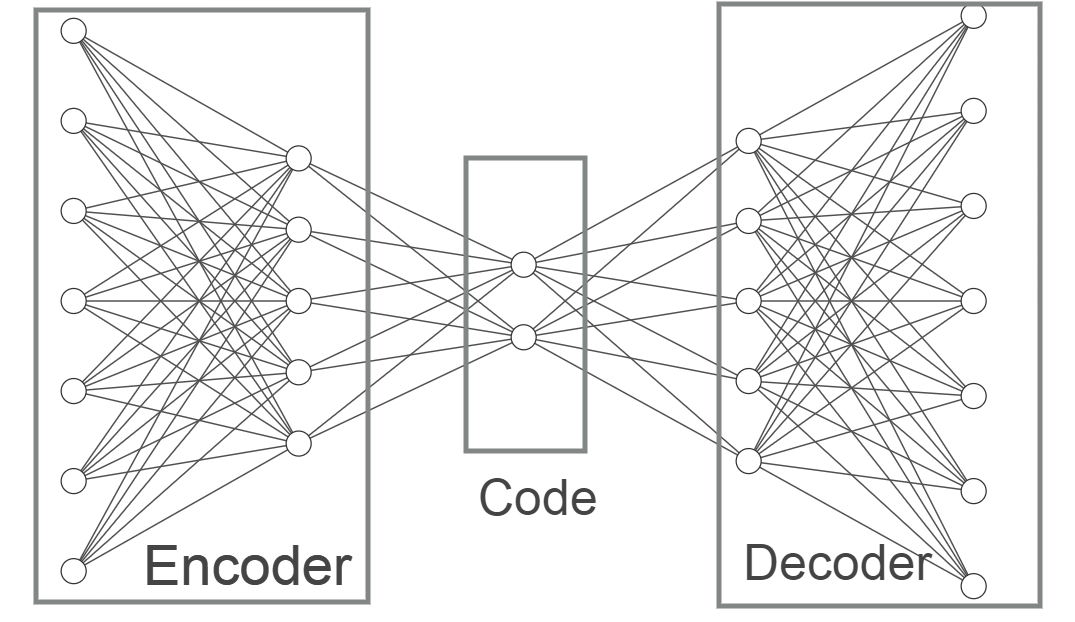}
	\caption{Simple schema of an  autoencoder with two code nodes.}
	\label{fig3}
\end{figure}
There are two major problems with using an autoencoder network:
\begin{enumerate}
  \item A typical  autoencoder network is symmetrical and has similar structures for encoder and decoder parts. This means code nodes generate outputs through a nonlinear transformation. 
  \item In a typical  autoencoder network there is no restriction on the correlation of code nodes. The code nodes could have a high correlation after calibrating the network. 
 
\end{enumerate}

In the following section, we propose a solution to address these issues.
\subsection{Variational Autoencoder}
\begin{figure}[H]
	\centering
	\includegraphics[width=0.8\textwidth]{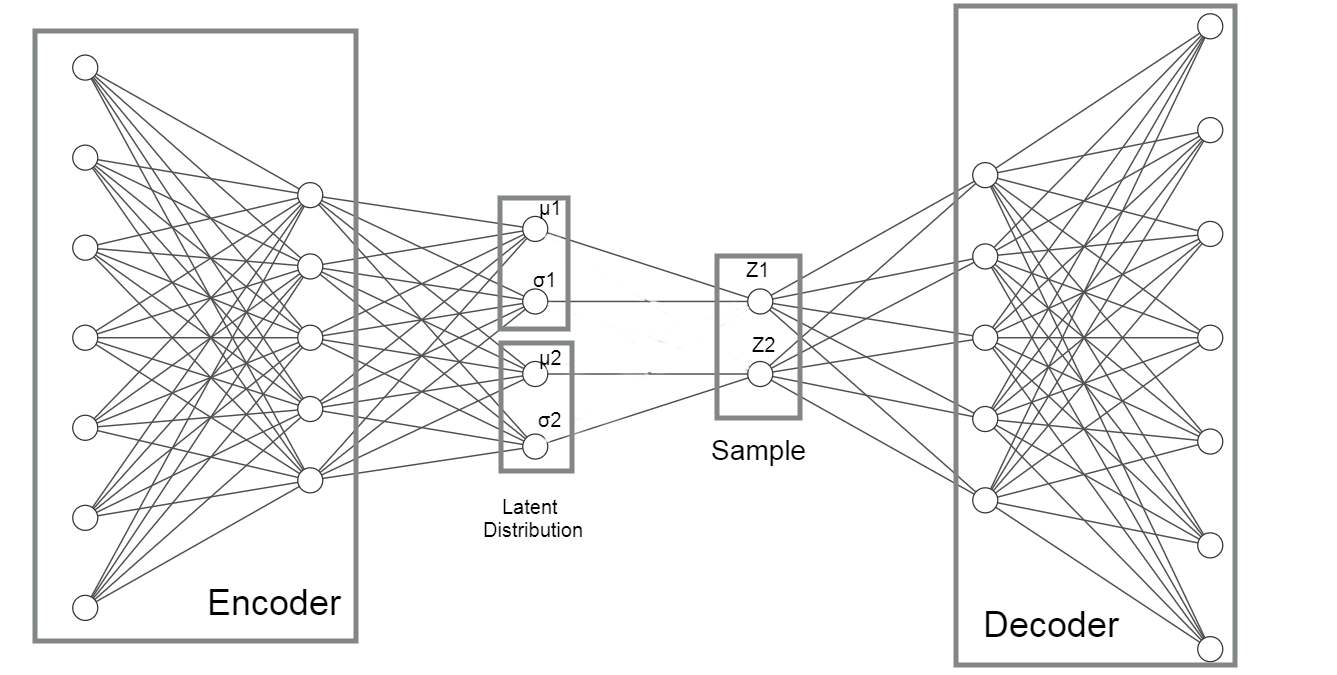}
	\caption{Simple schema of a  variational autoencoder with two latents.}
	\label{fig4}
\end{figure}

"In machine learning, a variational autoencoder (VAE), is an artificial neural network architecture introduced by Diederik P. Kingma and Max Welling, belonging to the families of probabilistic graphical models and variational Bayesian methods." \cite{aVar-encod}\\
In VAE, the dimension is reduced by mapping the input into a few independent standard normal distributed variables e.g. Z1 and Z2 in Figure \ref{fig4}. Then the encoder network maps back the Z values into the input variables.  The loss function in VAE is designed to not only makes the input close to the output but also  make Z values as close as possible to independent normal distributions by calibrating $\mu$ and $\sigma$ values. This will ensure the conditional independence of factors i.e. Z values. But the factors are mapped back into variables using nonlinear transformations which is not acceptable for our hedge ratios estimation problem. To address the issues we are proposing a Modified Variational Autoencoder network as we explain in the following section.\\

\subsection{Modified Beta Variational Autoencoder}
To explain the model, assume we want to hedge PnL of the main portfolio where its time series is presented by $Y(t)$. For the purpose of illustration, in this example, we assume to have two candidate hedging instruments with time series $H_1(t)$ and $H_2(t)$. 
\begin{eqnarray}
	Y(t) =   \beta_1 H_1(t) +\beta_2 H_2(t) + \epsilon(t)      
	\label{Eq22}
\end{eqnarray}

where ($E(\epsilon(t))=0$) i.e. the expected value of residuals is zero. Please note in general we are not expecting any intercepts. If there is any intercept, that means there is a market inefficiency to make money by going long the portfolio and shorting the hedging portfolio if the intercept is positive, and vice versa, if the intercept is negative. This migth exists in a short period of time but it will not last and we are not expecting it to be observed in the future. First, we construct the following neural network:\\

\begin{figure}[H]
	\centering
	\includegraphics[width=0.8\textwidth]{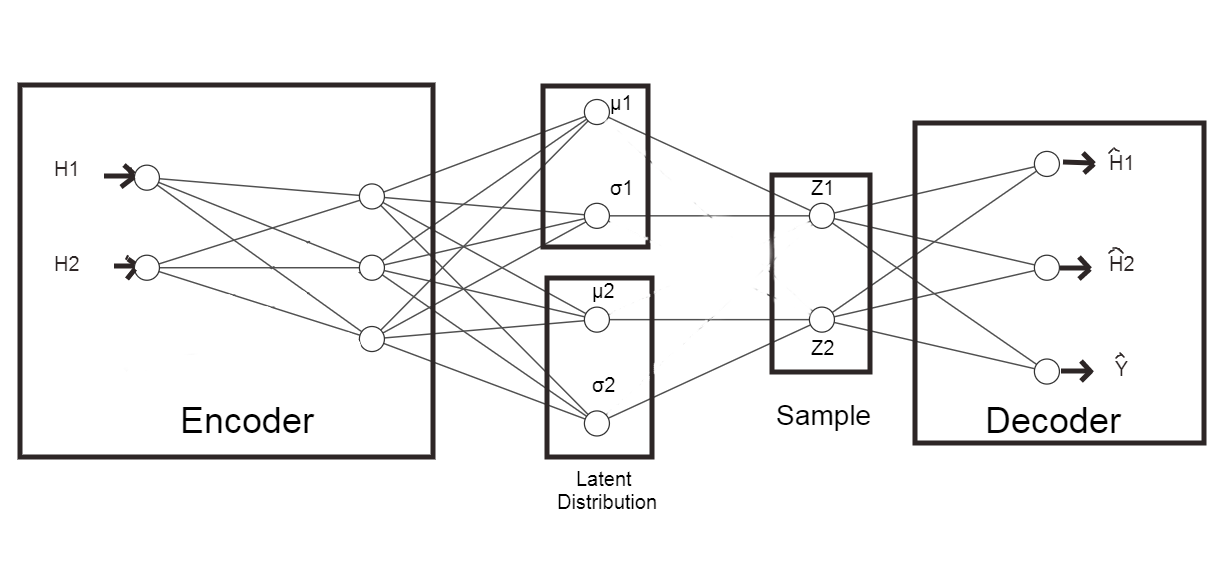}
	\caption{Modified variational autoencoder with two latents.}
	\label{fig5}
\end{figure}

There are several differences and similarities between the modified VAE and the original VAE. The loss function in both  minimizes not only the difference between targeted output values and estimated ones but also  pushes $\mu$ values to zero and $\sigma$ values to one.
\begin{eqnarray}
\begin{split}
	Loss(t) =  (H_1(t)-\hat{H_1}(t))^2 +(H_2(t)-\hat{H_2}(t))^2   +(Y(t)-\hat{Y}(t))^2  + \\ \hat{\beta}~( KL(N(\mu_1, \sigma_1, N(0,1))+KL(N(\mu_2, \sigma_2, N(0,1)))
	\label{Eq23}
 \end{split}
\end{eqnarray}
where $\hat{\beta}$ is a free parameter and KL is Kullback–Leibler divergence \cite{KL}. The use of $\hat{\beta}$ was introduced by Higgins et al. where $\hat{\beta}$ is not equal to one, the model is called Beta-VAE\cite{BVAE}. 
Please  note that:
\begin{itemize}
  \item Similar to $\beta$-VAE, modified $\beta$-VAE has the same structure i.e. encoder and decoder as well as latent distribution unit.
  \item While the encoder in modified $\beta$-VAE could have any number of hidden layers and activation functions, the number of inputs is equal to the number of hedge instruments.
  \item the decoder has the same number of outputs as the encoder plus one for Y.  It only has one layer with linear activation with no bias. 
  \item The latent distribution has twice the number of hedge instruments to model $\mu$ and $\sigma$ values.
\end{itemize}

The Z values are computed via a  non-linear  transformation of inputs while the outputs are computed from Z values by a linear transformation. Similar to linear factors, we can write the following equations corresponding to Figure \ref{fig5}:
\begin{eqnarray}
	    \bm{\hat{H}_{k \times 2}=Z_{k \times 2}~\gamma_{2 \times 2} }  
	\label{Eq24}
\end{eqnarray}
The value of matrix $\gamma_{2 \times 2}$ is extracted from the calibrated weight of the decoder. 
\begin{eqnarray}
	    \bm{\hat{Y}_{k \times 1}=Z_{k \times 2}~\alpha_{2 \times 1} }  
	\label{Eq25}
\end{eqnarray}
The value of matrix $\alpha_{2 \times 1}$ is extracted from the calibrated weight of the decoder to compute Y from Z values. Using \ref{Eq20}  we can compute the hedge ratios:
\begin{eqnarray}
	    \bm{\beta_{2 \times 1} = \gamma_{2 \times 2}^{-1}~\alpha_{2 \times 1}}  
	\label{Eq26}
\end{eqnarray}

\section{Historical values sampling}\label{Sec:Decay}
In all methods, we are using dependent i.e. $Y(t)$ and independent variables i.e. $X_{i}(t)$ variables. Typically we use historical values for Xs and Y time series. In reality, we are interested in the next move of the market rather than historical values.  We like to create a hedged portfolio so that the move in our position gets offset by the move in the hedge instruments so that the hedge position shows minimal PnL movement. Nobody knows what the next movements will be, but we can estimate the next movement distribution by considering historical movements. For example, we can assume that the next move will be similar to one of the movements observed in the last 250 days. By using the last 250 days' Xs and Y time series we are implicitly assuming the next-day movement distribution is the same distribution that has generated the last 250 samples. In this section, we present a tool on how to test this assumption and how to generate better samples to be used in our models for dependent and independent values.\\
Assume we have a time series X where we have observed the values till  the time "t". We would like to estimate X(t+1) distribution  given time series X:
\begin{eqnarray}
	    P(x(t+1)  ~|~ \bm{X} )= f(t+1)\\
            \bm{X} = [X(1),X(2),...,X(t)]
	\label{Eq27}
\end{eqnarray}
We can compute the quantile of x(t+1) assuming its distribution is f(t+1). Let's call that as $F^{-1} (x(t+1))$. We know if we have estimated f(.) correctly then $F^{-1} (.)$ should follow a uniform distribution regardless of what kind of distribution f(.) follows. For most of the financial time series, we could see that using historical value will not satisfy this condition. Intuitively we also expect to see the next day move more similar to today's move than similar to a move that happened 6 months ago. One way to address this concern is to generate Xs and Y time series by sampling from historical values with an exponential distribution where we assign a higher probability to more recent data to be picked than older data. 
Let's start with a univariate time series:
\begin{eqnarray}
	\bm{X} = [X(1),X(2),...,X(t)]\\
        \bm{P} = [\alpha^{t-1},,\alpha^{t-2},...,\alpha,1]\times G\\
        G= \sum_{i=1}^{i=t} \alpha^{t-i}
	\label{Eq29}
\end{eqnarray}
where G is the scaling factor. $\alpha \le 1.0$ is the decay factor where it could be changed so that $F^{-1} (.)$  become a uniformly distributed variable. \\
For multivariate time series, we can compute the principal component time series first which by construction are perpendicular i.e. they have zero correlation. Now we can treat each principal component time series as a univariate time series and compute its $\alpha$. The $\alpha$ for the whole set is the weighted average of each $\alpha$ computed for each principal component time series. The weights could be the percentage of total variance that each component explains in PCA. Please note that in multivariate time series, we sample dates first and then generate the whole matrix. In other words, we are sampling vectors of data (Y and X values)  from different dates to preserve the correlation.

\section{Result and Discussion }
In this section, we are illustrating  how different methods could be used. Assume we want to hedge (or compress risk) SPY using a few equities as hedging instruments (or risk factors). SPY (SPDR S\&P 500 ETF Trust) is an exchange-traded fund that trades on the NYSE Arca and tries to track the S\&P 500 stock market index. S\&P 500 is an equity basket tracking the stock performance of 500 large companies listed on stock exchanges in the United States. We use daily historical prices from 2014-03-27 to 2022-12-23. The following figures show SPY and candidate hedging instruments' relative price movements:\\
\begin{figure}[H]
	\centering	\includegraphics[width=0.65\textwidth]{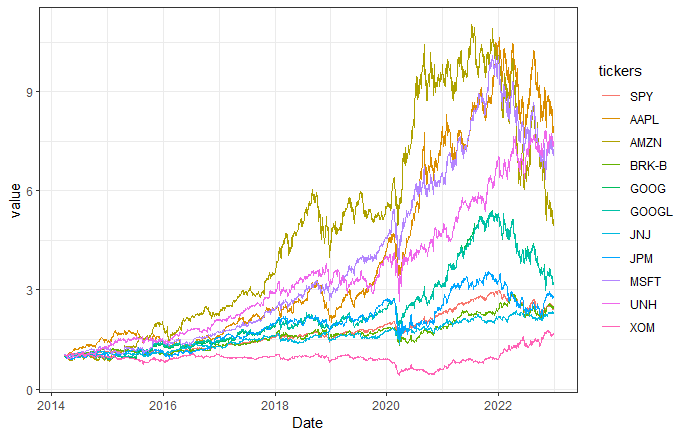}
	\caption{Relative price movement of SPY and candidate hedging instruments.}
	\label{fig6}
\end{figure}
\begin{figure}[H]
	\centering	\includegraphics[width=0.65\textwidth]{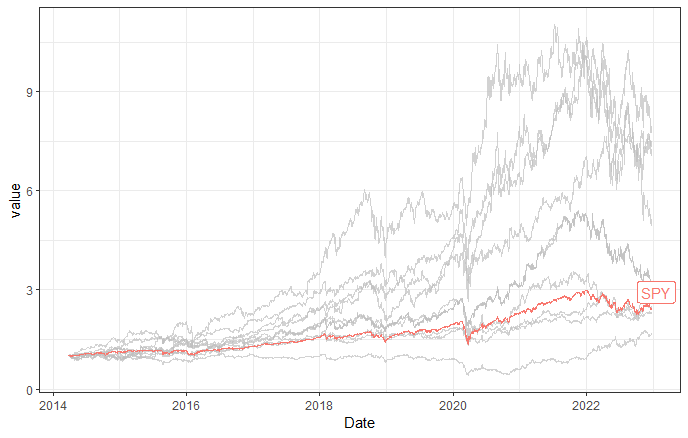}
	\caption{Relative price movement of SPY to other candidate instruments.}
	\label{fig7}
\end{figure}
The following figure  shows the correlation between the log rate of return of daily prices. As expected there are relatively high correlations between equities, especially between GOOG  and GOOGL tickers as expected:
\begin{figure}[H]
	\centering
	\includegraphics[width=0.75\textwidth]{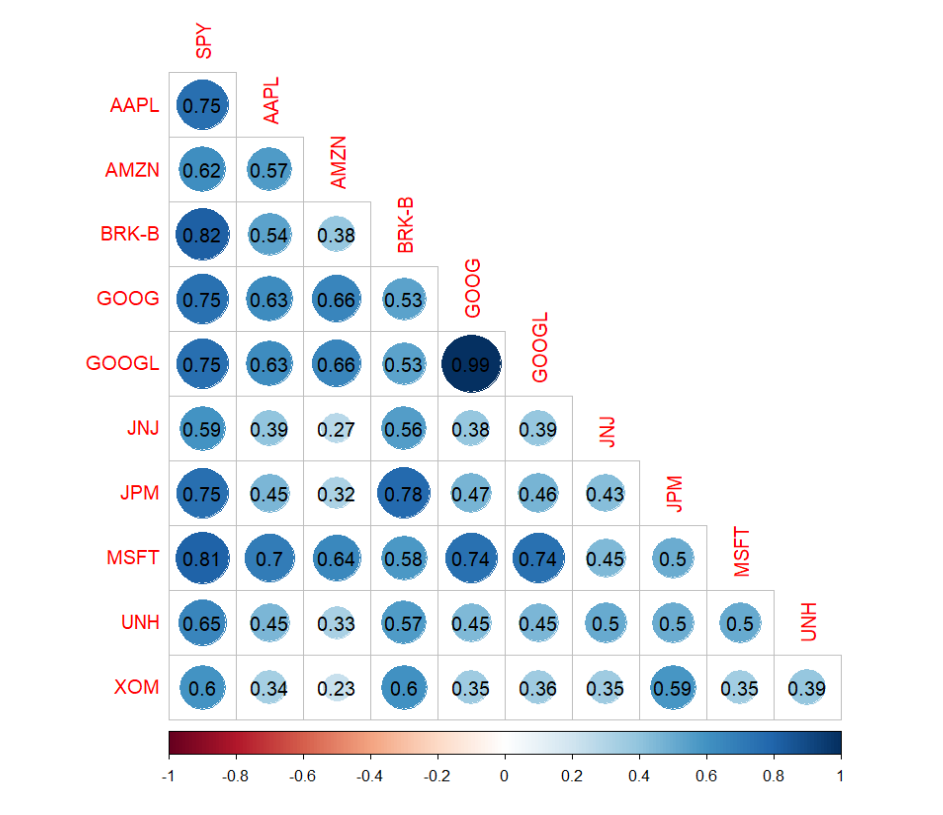}
	\caption{Correlation matrix. }
	\label{fig8}
\end{figure}
If we use different models as discussed in this paper, then we could compute the hedge ratios. The sampling decay as discussed in Section \ref{Sec:Decay} is equal to 0.996584 using the last 100 days. The following table shows the results for sampling 1000 data points:
\begin{table}[H]
\centering
\caption{Hedge ratios and R-Squared values for different methods}
	\label{HadgeRatios}
\resizebox{\textwidth}{!}{
\begin{tabular}{lcccccccc}
  \hline
Name & Regression & Lasso & Lasso\_Cost & Ridge & Ridge\_Cost & Factors\_PCA & Factors\_Varimax\_FA & VAE \\ 
  \hline
AAPL  & 0.135      & 0.136 & 0.147       & 0.136  & 0.136       & 0.136        & 0.113                  & -0.086 \\
AMZN  & 0.089      & 0.089 & 0.092       & 0.089  & 0.089       & 0.091        & 0.071                  & -0.501 \\
BRK-B & 0.251      & 0.248 & 0.191       & 0.243  & 0.244       & 0.249        & 0.101                  & 0.618  \\
GOOG  & -0.040     & 0.000 & 0.045       & -0.001 & -0.025      & -0.042       & 0.068                  & -3.462 \\
GOOGL & 0.068      & 0.027 & 0.000       & 0.031  & 0.054       & 0.070        & 0.070                  & 3.829  \\
JNJ   & 0.092      & 0.090 & 0.090       & 0.091  & 0.093       & 0.090        & 0.087                  & -0.093 \\
JPM   & 0.149      & 0.150 & 0.180       & 0.152  & 0.153       & 0.148        & 0.136                  & -0.336 \\
MSFT  & 0.126      & 0.126 & 0.106       & 0.124  & 0.125       & 0.124        & 0.097                  & 0.894  \\
UNH   & 0.015      & 0.014 & 0.009       & 0.016  & 0.015       & 0.017        & 0.133                  & 0.205  \\
XOM   & 0.063      & 0.063 & 0.066       & 0.064  & 0.064       & 0.063        & 0.095                  & 0.101  \\
   \hline
R2    & 0.958      & 0.958 & 0.957       & 0.958  & 0.958       & 0.958        & 0.943                  & 0.936 \\
   \hline
\end{tabular}
}
\end{table}
Columns represent employed methods and all rows except the last one are the computed hedge ratios to hedge SPY. The last row presents R squared which is computed as the squared correlation between the predicted values and actual values of the daily log return of SPY. For all methods, we consider the cost of hedging as shown in Table \ref{HadgeCost}. They have been computed according to equations \ref{Eq12} and \ref{Eq13}. As discussed before only Lasso\_Cost and Ridge\_Cost models actively consider the cost of hedges in optimization and finding hedge ratios. However, the cost of hedges is considered in the expected PnL as can be seen in Figure \ref{fig:boxplot} where the mean of each boxplot is equal to the cost of hedging rather than zero.

\begin{table}[H]
\centering
\caption{Hedging cost}
	\label{HadgeCost}

\begin{tabular}{lc}
  \hline
Variable & Cost     \\
  \hline
AAPL     & 0.000529 \\
AMZN     & 0.000082 \\
BRK-B    & 0.000863 \\
GOOG     & 0.000282 \\
GOOGL    & 0.000387 \\
JNJ      & 0.000304 \\
JPM      & 0.000252 \\
MSFT     & 0.000633 \\
UNH      & 0.000364 \\
XOM      & 0.000269\\
  \hline
\end{tabular}
\end{table}

The following figure shows the boxplot of residuals plus hedging cost and  99\% Value at Risk (VAR). The mean is equal to the cost of hedging which is negative. VaR is a function of  both the deterministic cost of hedging as well as the tail risk which is stochastic.  
\begin{figure}[H]
	\centering
	\includegraphics[width=0.75\textwidth]{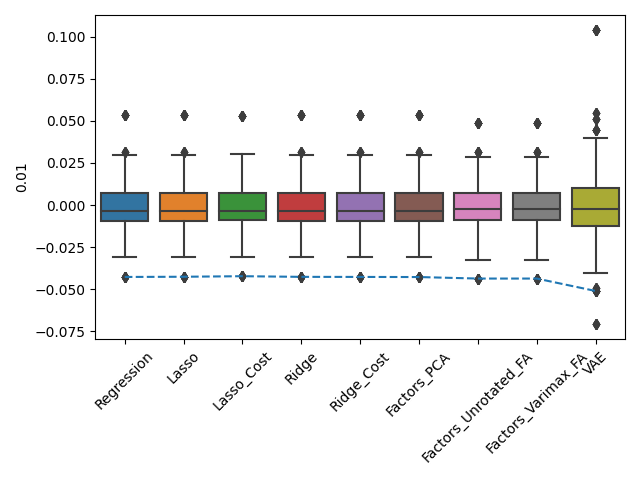}
	\caption{Different models residuals and 99\% Value at Risk (VAR), presented as the dashed line.  }
	\label{fig:boxplot}
\end{figure}

\subsection{Models Advantages and Limitations}{
Each model has its own advantages and limitations. Here we briefly discuss each model's pros and cons:
\begin{itemize}
  \item \textbf{Regression:} Regression is easy to perform and least computationally expensive. Given there is no restriction on coefficients it should provide the lowest in-sample fitted error variance. However, a regression could be prone to over-fitting and higher cost of hedging given there is no restriction on computed hedge ratios. It could also become unstable when highly correlated hedge instruments (independent variables) are used.
  \item \textbf{Lasso:} Lasso is similar to regression when L1 regularization is applied to  fitted coefficients i.e. hedge ratios. The regularization might force a number of coefficients to be set to zero and control the size of other coefficients. While regularization helps with avoiding over-fitting and controls the cost of hedging, we might get higher  in-sample fitted error variance given the applied restriction on the optimization. In practice, we can use Lasso as a feature engineering tool to pick a handful of hedge instruments from a pool of candidate hedge instruments. One of the limitations of the model is that model picks which coefficients are forced to zero as data changes. If for example, we use the tool for risk compression, it would be difficult to monitor coefficients over time given that any variable could be dropped out of the equation at any time. This also could increase the cost of hedge portfolio rebalancing.

  \item \textbf{Lasso\_Cost:} Lasso\_Cost is the same as Lasso where the cost of hedging is the same for all hedge instruments. Therefore, this model has the same advantages and limitations as Lasso. In addition, the model could adjust the computed weights to consider the relative costs between hedge instruments.  

\item \textbf{Ridge:} Ridge is similar to regression when L2 regularization is applied to  fitted coefficients i.e. hedge ratios. The regularization  forces the coefficients to be set to small values but not zero. While regularization helps with avoiding over-fitting and controls the cost of hedging, we might get higher  in-sample fitted error variance given the applied restriction on the optimization similar to Lasso. In practice, we can use Ridge when we want to keep all variables in contrast to Lasso where we need to select a subset of variables. 

\item \textbf{Ridge\_Cost:} Ridge\_Cost is the same as Ridge where the cost of hedging is the same for all hedge instruments. Therefore, this model has the same advantages and limitations as Ridge. In addition, the model could adjust the computed weights to consider the relative costs between hedge instruments. 

\item \textbf{Factors\_PCA:} In this method we first compute the PCA time series as factors. Using PCA by design ensures the factors to be perpendicular. Using Factors, in general, are more intuitive to understand since we find hedge ratios so that we do not have any exposure to factors. However, given it is a two-step process i.e. finding factors under some restrictions and then finding hedge ratios that eliminate our exposures to factors, we are expecting, in general,  to have a less optimum solution given more restrictions in the process.

\item \textbf{Factors\_Unrotated\_FA:} In this method we first compute the factors' time series using factor analysis \cite{factor}. The pros and cons of this method are similar to Factors\_PCA.

\item \textbf{Factors\_Varimax\_FA:} As it could be shown, extracted factors using factor analysis \cite{factor} are not unique and rotated factors are also a possible answer. "Varimax is the most commonly used rotation method. Varimax is an orthogonal rotation of the factor axes that maximizes the variance of the squared loadings of a factor (column) on all the variables (rows) in a factor loadings matrix"  \cite{varimax}. Please note that the rotation only helps with better interoperability of the results i.e. which factors explain what variables. The final results for the hedge ratios are the same for Factors\_Varimax\_FA and Factors\_Unrotated\_FA methods.  The pros and cons of this method is similar to Factors\_PCA.
\item \textbf{VAE:}  This  model is a modified $\beta-$VAE. The model is similar to a common factor analysis. The main difference is that factors are generated non-linearly while in PCA and Factor Analysis, the factors are constructed from variables linearly. In addition, VAE is a neural network where it inherited the  pros and cons of the neural network models. One of the limitations is that VAE is  sensitive to the initial seed for its weights initiation in the optimization and instability in the results from one run to the other \cite{nnt_issues}. Training the model might take longer than other models and we might  not observe better performance than other models given higher computation cost in general.  
  
\end{itemize}

}
\section{Conclusion}\label{Sec:conc}
Besides a simple linear regression, there are other methods to use to find hedge ratios or compress risk in a few risk factors. The methods fall into either Regularization Techniques or Common Factor Analyses. In general Regularization Techniques  could not only help with the overfitting issues but also control the size of the hedges to save hedging costs. Lasso which is a member of this family   could  be used to select hedge instruments from a pool of candidate securities. In regularization methods, we could also consider the cost of hedging in the optimization which is a function of the type of security, market volatility, and the liquidity of the hedging instruments. \\
In Common Factor Analyses methods the securities first map into common factors and then hedges are selected in a way to kill the exposure of the hedged portfolio to those factors. While this method might provide a more intuitive solution, it might not provide the best answers given the two-step process in comparison to Regularization Techniques. We also can't control the size of the hedges similar to the Regularization Technique. However, both Regularization Techniques and Common Factor Analyses are robust to deal with multicollinearity. This means we can use highly correlated hedge securities and find their hedge ratios  without generating instability that could be observed by using a linear regression model. We presented a modified beta variational autoencoder model that could be used for hedging. This method falls into Common Factor Analyses type methods. \\
We have shown using historical values without proper sampling might not be the best option to calibrate our models. We  discussed and presents a method on how to sample from the historical values to be used as input data in our models. In this method, we calibrate an exponential distribution where it samples from historical values by giving more weight to more recent data than older data.  \\
Finally, numerical results have been generated for an example using different methods for comparison. As discussed in section \ref{Sec:conc},  each method has its pros and cons and a proper method should be used for each problem. As part of the assessment not only the variation of the hedged portfolio PnL minimization but also the PnL associated with the cost of hedging should be considered.


\bibliographystyle{IEEEtran}



\end{document}